\documentclass[final,5p,times,twocolumn]{elsarticle}

\usepackage{lineno,hyperref,amsmath}
\usepackage{xcolor}
\modulolinenumbers[200]

\graphicspath{ {figures/} }

\journal{High Energy Density Physics}

\begin{document}

\begin{frontmatter}

\title{The STAG Code: A Fully Relativistic Super Transition Array Calculation Using Green's Functions}

\author[lanl]{N. M. Gill}
\ead{ngill@lanl.gov}

\author[lanl]{C. J. Fontes}
\author[lanl]{C. E. Starrett}

\address[lanl]{Los Alamos National Laboratory, P.O. Box 1663, Los Alamos, NM, 87545, USA}

\begin{abstract}
Calculating opacities for a wide range of plasma conditions (i.e. temperature, density, element) requires detailed knowledge of the plasma configuration space and electronic structure. For plasmas composed of heavier elements, relativistic effects are important in both the electronic structure and the details of opacity spectra. We extend our previously described superconfiguration and super transition array capabilities [N. M. Gill et al., JPB, 56, 015001 (2023)] to include a fully relativistic formalism. The use of hybrid bound-continuum supershells in our superconfigurations demonstrates the importance of a consistent treatment of bound and continuum electrons in dense plasma opacities, and we expand the discussion of these consequences to include issues associated with equation of state and electron correlations between bound and continuum electrons.
\end{abstract}


\end{frontmatter}

\section{Introduction}
Understanding the interaction of light with plasma ions is important for understanding and simulating the dynamics of many systems, including stellar interiors~\cite{serenelli2009,salmon2012,piron2018}, inertial confinement fusion implosions~\cite{nuckolls1972,hu2018}, and high-energy-density experiments, such as those performed at the Sandia National Laboratories' Z-facility~\cite{bailey2007,bailey2015}. One of the most important quantities needed to model such interactions is the radiative opacity, which is a measure of the probability that light of certain energies will be absorbed by plasmas. Recent discrepancies between theoretically predicted and experimentally measured opacities has motivated the search for improved theoretical modeling of plasmas~\cite{bailey2015,colgan2016, mancini2016, more2017, NAGAYAMA201617, nagayama2019, gill2019,shaffer2022,gill2021,zeng2021,Mondet_2015,delahaye2021}. 

When a plasma is modeled as a system of independent ions, calculating the opacity requires knowledge of all possible atomic configurations in which each ion can exist. Describing the statistical properties of a plasma is done through the use of thermodynamic state variables, from which one can obtain the populations of each atomic configuration (i.e. the probability that a plasma ion would be found in a state described by a given atomic configuration). The configuration populations, along with the photo-absorption cross-sections associated with each configuration, are used to construct the opacity spectrum of the entire plasma.

In order to address the fact that the plasma environment (e.g. the presence of continuum electrons, collisions and correlations between ions, etc.) influences the atomic structure of atomic configurations, quantum average atom models (AA) were developed to describe the average electronic structure of plasma ions~\cite{liberman,rozsnyai,mermin65}. Most AA models rely on the so-called ion-sphere approach, which treats the ions as spherical cavities in which the electrons are distributed among independent bound and continuum quantum states, and generally the occupation of these states is governed by the Fermi statistics associated with a self-consistently calculated thermodynamic state of the plasma (i.e. the electronic states, occupations of those states, and the thermodynamic state variables are calculated self-consistently). 

With crude but implicit inclusion of plasma effects, AA calculations allow for a description of plasma electronic structure in an average way with only a single calculation of atomic structure. Representing all atomic configurations with a single, averaged electronic structure has led to significant successes when determining thermally averaged quantities such as the equation of state~\cite{liberman, blenski, wilson06, sterne07, starrett2018}, but the spectral opacity generated from a single, average electronic structure lacks the necessary spectral details to be of use in quantitative comparisons~\cite{shaffer2022,gill2021,Bar-Shalom89}. 

In order to utilize the benefits of averaging over many configurations while still being able to use the averaged electronic structures to generate reasonable opacity spectra, Bar-Shalom and colleagues developed the concept of superconfigurations~(SCs) and Super Transition Arrays~(STAs)~\cite{Bar-Shalom89}. SCs are defined simply as groups of configurations, and such a grouping can range from including all relevant configurations in a single SC (which is equivalent to the AA model) down to each SC containing only a single configuration (which is equivalent to the so-called configuration-average model). Each SC has an associated electronic structure that approximates the thermally averaged electronic structure of all constituent configurations. The STA formalism allows one to approximate the statistical properties associated with the spectral response of all constituent configurations within a given SC (i.e. the statistical moments of the configuration-average spectrum are approximated by the STA formalism). The combination of carefully defined SCs and STA techniques allows one to obtain opacity spectra that compare well with approaches that require an explicit list of all relevant configurations~\cite{Bar-Shalom89,bar-shalom96,BLENSKI2000,porcherot_scorcg_2011,krief2018,opus2020,Gill_2023}.

In this work, we report on developments to the previously described SC and STA code developed at Los Alamos National Laboratory~\cite{Gill_2023}. That previously described capability utilized the novel formalism of hybrid bound-continuum supershells, which incorporate the continuum electronic structure on a consistent theoretical basis with the bound electronic states by utilizing a Green's function approach to calculate the electronic structure of the hybrid supershells. Recent developments have extended the code's capabilities to include a fully relativistic treatment of SC electronic structure and the spectral response of said structure. The fully relativistic and non-relativistic capabilities are integrated into the Super Transition Array using Green's functions (\texttt{STAG}) code. 

In section~2, we briefly describe the extension of the formalism outlined in reference~\cite{Gill_2023} to include relativistic effects. In section~3, we provide results that demonstrate the differences between relativistic and non-relativistic STA spectra for an iron plasma as a validation of the code functionality. In section~4, we reiterate the previously shown effects of a consistent treatment of bound and continuum states and discuss the additional physics that should be included to improve this consistency within the opacity formalism.

\section{Theoretical Description}
A relativistic atomic configuration is defined by the occupation of independent electronic eigenstates of an effective Hamilton. These eigenstates can be characterized by the principal quantum number ($n$), the azimuthal quantum number ($l$), and the total angular momentum number quantum number ($j=l\pm\frac{1}{2}$). Each unique set of $nlj$ quantum numbers is often called an orbital, which has its own energy and wavefunction. The occupations of these orbitals by an integer number of electrons is what uniquely defines each atomic configuration ($C$), i.e.
\begin{equation}
\label{config_def}
C = \prod_{s} (n_{s} l_{s} j_{s})^{q_{s}} \,,
\end{equation}
where the product runs over all possible orbitals and $q_{s}$ is the occupation of orbital $s$, which can take integer values ranging from $0$ to the statistical weight of the orbital, $g_{s}=2j+1$.

An SC ($\Xi$) is a group of configurations~\cite{Bar-Shalom89}, and we define SCs by the occupation of so-called supershells, which are groups of orbitals, i.e. 
\begin{equation}
\label{sc_def}
\Xi = \prod_{\sigma} \Bigg(\prod_{s \in s_{\sigma}} n_{s}l_{s}j_{s}\Bigg)^{q_{\sigma}} =\prod_{\sigma} \sigma^{q_{\sigma}} \,,
\end{equation}
where $\sigma$ denotes a supershell, $q_{\sigma}$ is the integer occupation number of each supershell, and $s_{\sigma}$ is the set of orbitals contained within each supershell. The definition of a supershell, denoted by a particular set of orbitals $s_{\sigma}$, is typically chosen such that the difference in orbital energy between the most deeply bound and least deeply bound orbital in the supershell is less than the thermal energy of the system, $k_{B}T$.

Our method also includes the use of a hybrid bound-continuum supershell to encompass some negative energy orbitals near the positive energy continuum, as well as all of the continuum orbitals. This so-called continuum supershell allows us to smoothly capture the phenomenon of pressure ionization in our electronic structure calculations, where bound orbitals near the continuum edge are delocalized and reappear as continuum resonances due to plasma density effects.

Once an SC is defined, the electronic structure associated with that SC (i.e. the orbital energies and wavefunctions) is determined by solving the relativistic density functional theory (DFT) equations (for more information, see references~\cite{mermin65,kohn_sham_65,macdonald,gill17,starrett2018,Gill_2023}). The necessary components to solve these equations are the effective potential, $V_{eff}$, and the electron density, $n(\vec{r})$. The effective potential is defined in terms of the electron density as follows:
\begin{equation}
    \label{v_eff}
    V_{eff}(\vec{r}) = -\frac{Z}{r} + \int_{V_{ion}} d\vec{r'} \frac{n(\vec{r'})}{|\vec{r}-\vec{r'}|} + V^{xc}(\vec{r}) \,,
\end{equation}
where $Z$ is the nuclear charge, $V_{ion}$ is the volume of the ion-sphere, and $V^{xc}$ is the exchange-correlation potential, which is a functional of the electron density. 

The electron density of each SC is defined based on the contribution from each constituent supershell, i.e.
\begin{equation}
\label{edens_def}
n(\vec{r}) = \sum_{\sigma} n_{\sigma}(\vec{r}) + n_{c}(\vec{r}) \,,
\end{equation}
where the $n_{\sigma}$ are the electron densities associated with each bound supershell, and $n_{c}$ is the electron density associated with the hybrid continuum supershell.

The bound supershell electron densities are defined in terms of the orbital wavefunctions as
\begin{equation}
\label{ss_dens}
n_{\sigma}(r) = \sum_{s \in s_{\sigma}} f( E_{n_{s}l_{s}j_{s}}, \mu_{\sigma}) \, \frac{2j_{s}+1}{4\pi r^{2}}  \, \big[ \, P_{n_{s}l_{s}j_{s}}^{2}(r)+ Q_{n_{s}l_{s}j_{s}}^{2}(r)\,\big] \,,
\end{equation}
where the $P$ ($Q$) is the large (small) component of the radial wavefunction, $f$ is the Fermi-Dirac function, and $\mu_{\sigma}$ is an effective chemical potential that is varied until the following condition is met:
\begin{equation}
\label{ss_dens_charge}
q_{\sigma} = \int_{\, 0}^{\,R_{IS}} dr \, (4 \pi r^{2}) n_{\sigma}(r) \,.
\end{equation}

The electron density of the continuum supershell is defined by
\begin{equation}
\label{ctm_dens}
n_{c}(r) = \int_{E_{\rm min}}^{\infty} dE \, f(E,\mu) \, \sum_{l,j} \frac{2j+1}{4\pi r^{2}} \big[ \,P_{E,l,j}^{2}(r) + Q_{E,l,j}^{2}(r)\, \big] \,,
\end{equation}
where $E_{min}$ is the energy of the most deeply bound state included in the continuum supershell, and $\mu$ is the continuum chemical potential which is varied such that the following condition is met, i.e.
\begin{equation}
\label{ctm_dens_charge}
Z-\sum_{\sigma} q_{\sigma} = \int_{\, 0}^{\,R_{IS}} dr \, (4 \pi r^{2} ) n_{c}(r) \,.
\end{equation}

In practice, we utilize an equivalent formalism using Green's functions in order to obtain an alternate form of equation \ref{ctm_dens}, i.e.
\begin{equation}
\label{gf_c_dens}
n_{c}(r) = -\frac{1}{\pi} \Im \int_{E_{min}}^{\infty} dE \, f(E,\mu) \, Tr \, G(r,E) \,,
\end{equation}
where $G$ is the Green's function associated with the effective Hamiltonian. This alternate form allows us to carry out the integral over energies with a complex contour, which significantly improves the numerical efficiency and robustness of the calculation~\cite{gill17,starrett2018,Gill_2023}.

Equations~\ref{v_eff}--\ref{gf_c_dens} are solved iteratively until a converged electron density (and associated set of orbital energies and wavefunctions) is obtained. This iterative procedure requires an initial guess for $V_{eff}$, for which we use the $V_{eff}$ provided by the AA calculation from the \texttt{TARTARUS} code~\cite{gill17,starrett2018}.

Apart from the use of relativistic electronic structure, the form of the standard STA equations are unchanged from their non-relativistic form. We employ the standard STA formulas~\cite{Bar-Shalom89,bar-shalom96,Gill_2023} along with the efficient partition function routines of Gilleron and Pain~\cite{gilleron2004,wilson2007_part,pain2020_recursion}. 

\section{Comparisons Between Non-relativistic and Relativistic Calculations}
Beyond the improved physical description of atomic systems obtained within a relativistic electronic structure calculation~\cite{cowan1981theory} (such as through the inclusion of the spin-orbit interaction), we can see the importance of having a relativistic STA calculation when we compare the spectral resolution of a relativistic STA opacity with its non-relativistic counterpart. The capability to permute electrons between the different $nlj$ orbitals that arise from a given $nl$ orbital can significantly expand the size of the configuration space used to describe the plasma, and this larger space results in many more spectral lines between the increased number of configurations.

\begin{figure}[h]
    \centering
    \includegraphics[scale=0.33]{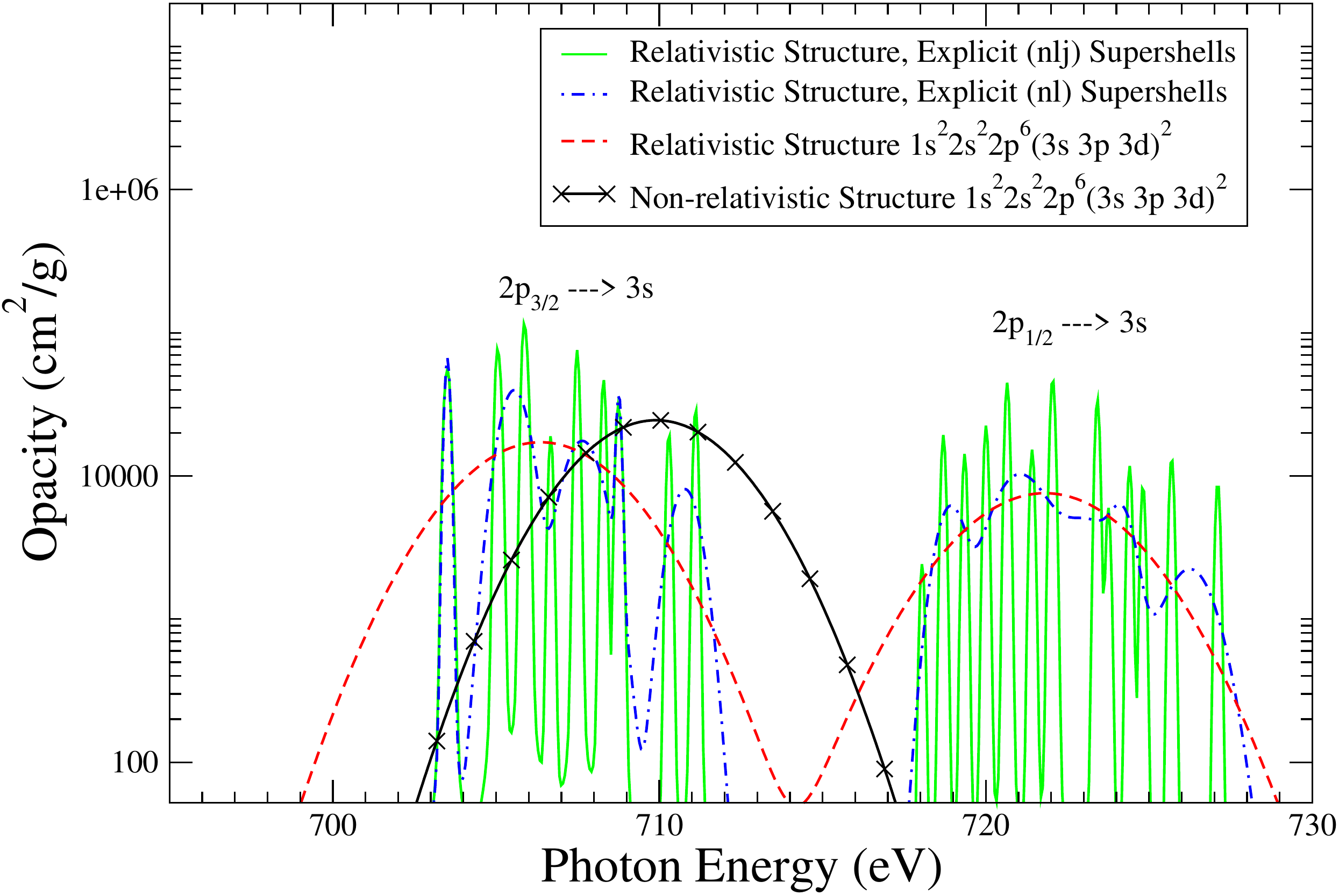}
    \caption[\fontsize{10pt}{10pt}]{A comparison of various bound-bound STA opacities is shown for an iron plasma at 180~eV and 0.1~$\textrm{g/cm}^{3}$, all of which are associated with the $2p_{3/2}~\rightarrow~3s$ and $2p_{1/2}~\rightarrow~3s$ transitions. The non-relativistic result is shown as solid black (with "$\times$" symbols) for the indicated SC, which appears as a single curve due to the degeneracy of the $2p_{3/2}$ and $2p_{1/2}$ orbitals. The red dashed curve shows the relativistic STA features associated with the same SC as before, but here the spin-orbit splitting leads to two distinct STAs. If the electrons in the coarse SC are permuted among $nl$-grouped supershells, we obtain the blue dash-dot curve, which is now associated with six distinct SCs. Permuting the electrons among $nlj$-grouped supershells produces an equivalent description to a configuration-average approach with 15 distinct relativistic configurations, which is represented by the solid green curve.}
    \label{fig:sta_rel_v_nrel}
\end{figure}

Figure~\ref{fig:sta_rel_v_nrel} contains a comparison of the bound-bound STA opacity for \hbox{$2p\rightarrow3s$} transitions arising from various SC definitions in an iron plasma at 180~eV and 0.1~$\textrm{g/cm}^{3}$. The black curve with ``$\times$" symbols represents the non-relativistic $2p \rightarrow 3s$ STA for transitions from the SC $1s^{2}\,2s^{2}\,2p^{6}\,(3s\,3p\,3d)^{2}$, and the dashed red curve represents the corresponding relativistic STAs for transitions from the equivalent relativistic SC (i.e. the $2p_{3/2}\rightarrow3s$ and $2p_{1/2}\rightarrow3s$ STAs associated with the $1s^{2}\,2s^{2}\,(2p_{1/2}\,2p_{3/2})^{6}\,(3s\,3p_{1/2}\,3p_{3/2}\,3d_{3/2}\,3d_{5/2})^{2}$ SC). The $2p_{3/2}~\rightarrow~3s$ and $2p_{1/2}~\rightarrow~3s$ relativistic STAs have significantly different energies, resulting in the two distinct features displayed by the dashed red curve in Figure~\ref{fig:sta_rel_v_nrel}. This energy splitting results from the inclusion of the spin-orbit
interaction in the Dirac equation. For elements lighter than iron, theory predicts that the two relativistic
energies will become progressively more degenerate as the atomic number decreases and the sum of the two relativistic STAs
will produce a single feature that is equal to the single non-relativistic STA.
For heavier elements, the energy separation will increase with atomic number and the two relativistic STAs will become even more distinct than what is displayed in the figure.

Further spectral resolution can be obtained by refining the SC through reduction of the supershell size. For example, if we define supershells according to an $nl$-grouped prescription, then significantly more spectral resolution can be obtained at the cost of increasing the number of explicit SCs required for our calculations. The dash-dot blue curve includes such a supershell definition, in which the $2p_{1/2}$ and $2p_{3/2}$ orbitals are included in a single supershell, with similar choices for the $3p_{1/2}$ and $3p_{3/2}$, and $3d_{3/2}$ and $3d_{5/2}$ orbital pairings. However, we note that this definition requires six explicit SCs
(e.g. $1s^{2}\,2s^{2}\,(2p_{1/2}\,2p_{3/2})^{6}\,(3s)^{0}\,(3p_{1/2}\,3p_{3/2})^{1}\,(3d_{3/2}\,3d_{5/2})^{1}$) to cover the same configuration space as the red curve covers with a single SC. From figure~\ref{fig:sta_rel_v_nrel}, we see that the STA energies and variance of the $nl$-grouped SCs do an excellent job of capturing the statistical properties of the configuration-average transition arrays (e.g. transition arrays arising due to transitions between configurations such as $1s^{2}\,2s^{2}\,(2p_{1/2})^{2}\,(2p_{3/2})^{4}\,(3s)^{0}\,(3p_{1/2})^{1}\,(3p_{3/2})^{0}\,(3d_{3/2})^{1}$ and $1s^{2}\,2s^{2}\,(2p_{1/2})^{2}\,(2p_{3/2})^{3}\,(3s)^{1}\,(3p_{1/2})^{1}\,(3p_{3/2})^{0}\,(3d_{3/2})^{1}$), which are shown in the solid green curve. It is worth noting here that the more averaged curves are centered nearest the strongest configuration-average lines, and this property of the STA energy demonstrates the power of capturing a strength-weighted average of the underlying transitions arrays with a single STA calculation. For more complicated SCs, in which the number of constituent configurations can easily number in the billions, a statistical approach such as the STA technique is the only means of capturing the influence of the many possible transition arrays.

\begin{figure}[h]
    \centering
    \includegraphics[scale=0.33]{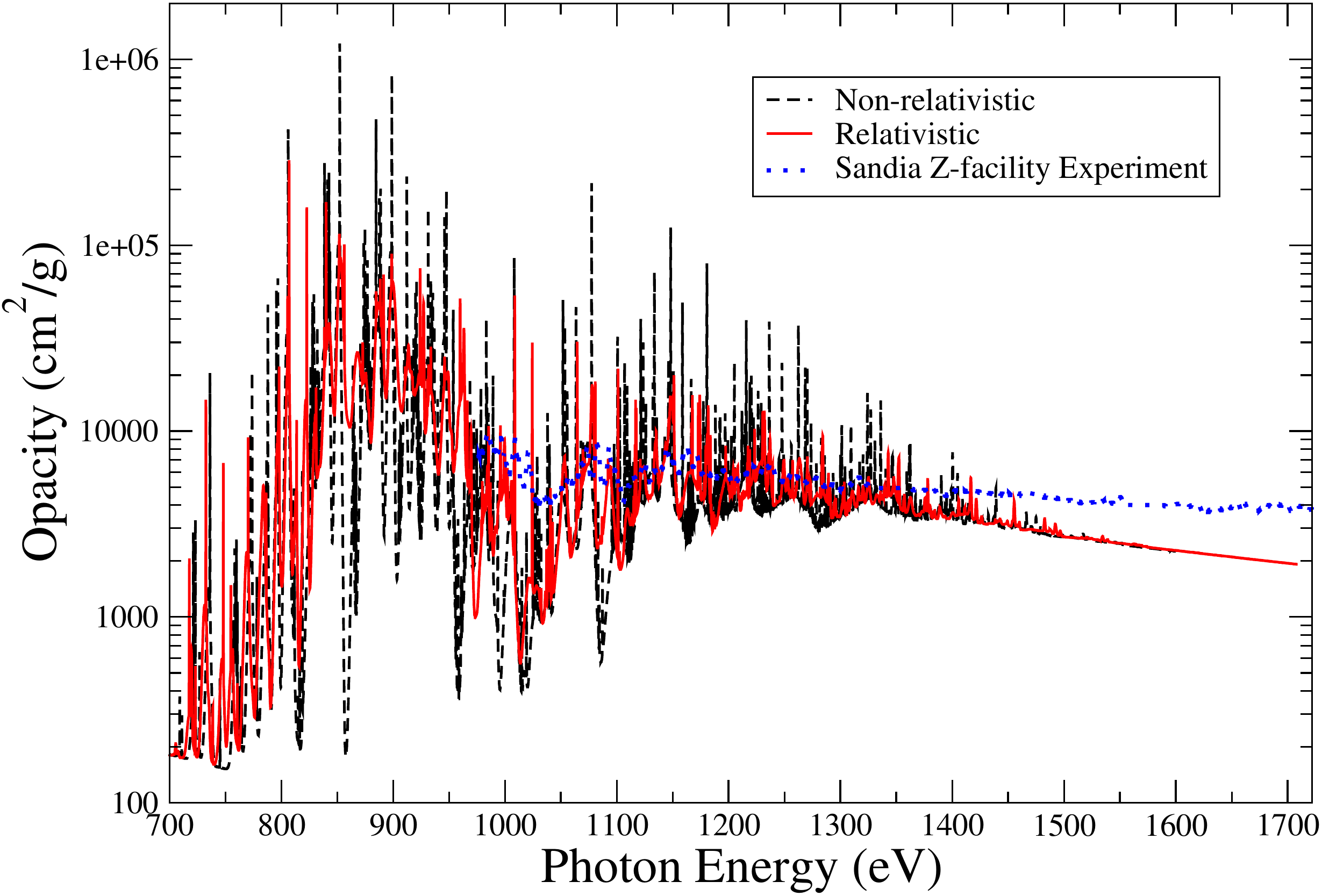}
    \caption[\fontsize{10pt}{10pt}]{The opacity for an iron plasma at 180~eV and 0.16~$\textrm{g/cm}^{3}$ is compared between non-relativistic and relativistic STA calculations. The black dashed curve shows the non-relativistic results, while the red solid curve shows relativistic results. Both calculations use the same definition for SCs, but the lack of degeneracy between $j$-resolved orbitals in the relativistic calculation leads to the opacity being distributed over many more lines than in the non-relativistic calculation. The results from the Sandia Z-facility experiments are shown in the blue dotted curve. The differences between the experimental and calculated lines are consistent with the observed discrepancies between the experiment and other model predictions.}
    \label{fig:fe_opac_rel_v_nrel}
\end{figure}

As a final demonstration of the differences between a non-relativistic and fully relativistic approach, we show the opacity for a complete description of an iron plasma at 180~eV and 0.16~$\textrm{g/cm}^{3}$ in figure~\ref{fig:fe_opac_rel_v_nrel}. These models use the same $nl$-grouped supershell definitions up to a valence supershell that contains all remaining orbitals, i.e. the SCs take the form $(1s)^{q_{1}}\,(2s)^{q_{2}}\,(2p)^{q_{3}}\,(3s)^{q_{4}}\,(3p)^{q_{5}}\,(3d)^{q_{6}}\,(4s\,4p\,4d\dots)^{q_{7}}$. We can see that the relativistic STAs in the solid red curve fill in many of the lower windows present in the non-relativistic STA curve in the dashed black curve. This filling is due to the splitting of $nl \rightarrow n'l'$ transition arrays into their $nlj \rightarrow n'l'j'$ counterparts, which leads to a distribution of the non-relativistic opacity over many more lines. In general, this phenomenon lowers the opacity peaks and raises the windows between clusters of lines in the relativistic calculation. 

The experimentally obtained results from the Sandia Z-facility for this case are shown as the blue dotted curve in figure~\ref{fig:fe_opac_rel_v_nrel} \cite{bailey2015,nagayama2019}. The discrepancy between model predictions and experiment that are observed in the bound-free opacity and in the windows between bound-bound features is consistent with other comparisons, e.g. References~\cite{bailey2015,nagayama2019,shaffer2022,gill2021,Gill_2023}. Further, discrepancies are present in the comparison due to the fact that only Doppler and natural broadening are included in the physical line-broadening of the \texttt{STAG} results shown in this work, whereas instrumental broadening as well as Stark and electron collisional broadening (which dominate the physical broadening at these conditions) are not present.

\section{Discussion of Continuum Influence on Opacity}
Our previous work~\cite{Gill_2023} has shown that the influence of the electronic structure associated with the continuum electrons can lead to significant differences in the opacity. In this section we will briefly discuss some of the importance of these differences and how a more consistent treatment of the continuum electrons in the opacity formalism may further improve descriptions of plasma properties.

\begin{figure}[h]
    \centering
    \includegraphics[scale=0.33]{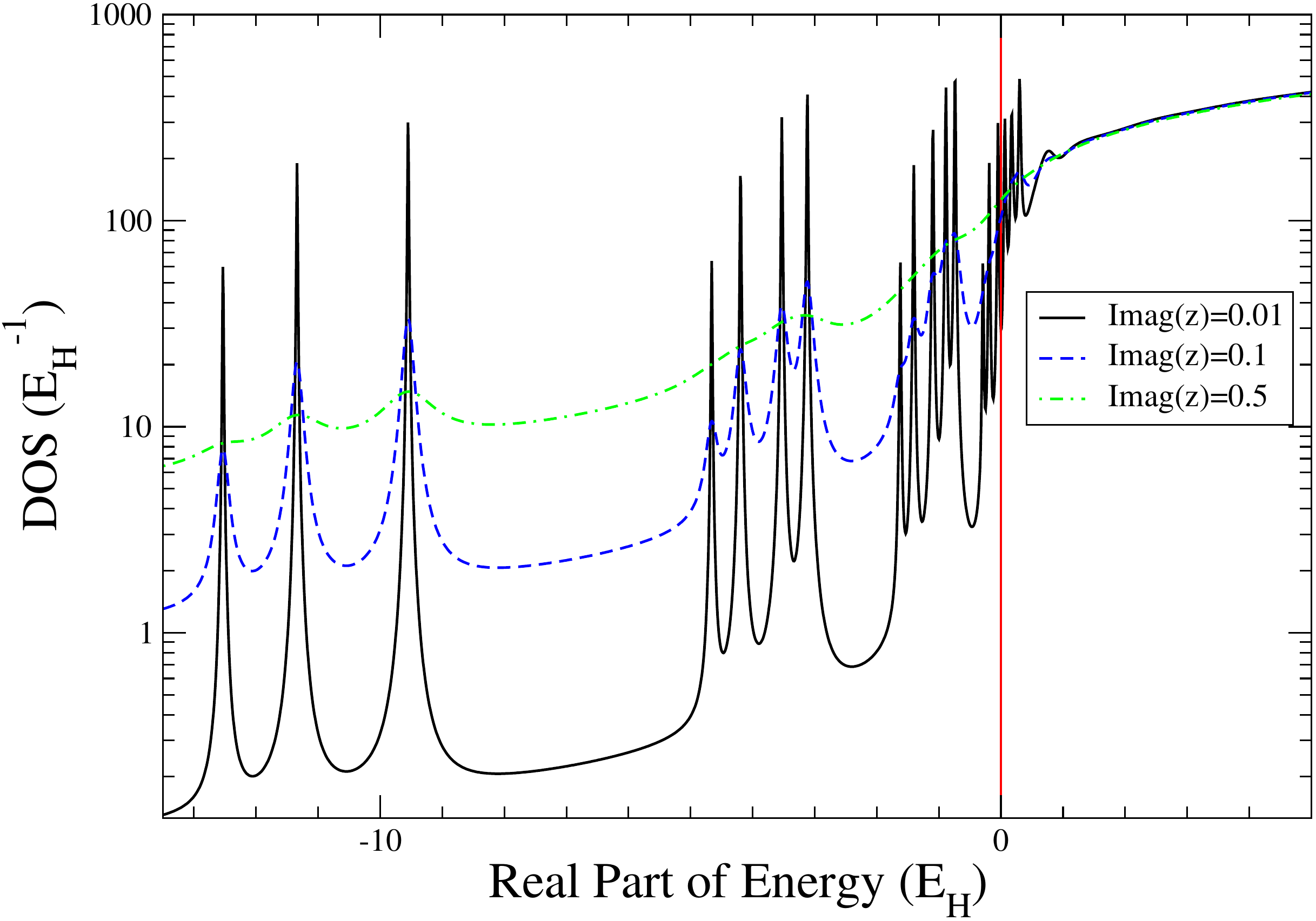}
    \caption[\fontsize{10pt}{10pt}]{The density of states along the top of a complex contour for an AA electronic structure of an iron plasma at 120~eV and 0.5~$\textrm{g/cm}^{3}$ is shown. The black solid curve is the DOS only 0.01~$\textrm{E}_{\textrm{H}}$ off of the real-energy axis, which results in a small broadening of the features as they would appear on the real-energy axis. The blue dashed curve is 0.1~$\textrm{E}_{\textrm{H}}$ off of the real-energy axis, and the sharp features have mostly been broadened into smooth features, especially near the continuum edge. The green dash-dot curve is 0.5~$\textrm{E}_{\textrm{H}}$ off of the real-energy axis, and this greatly broadened curve can be numerically resolved with very few grid points.}
    \label{fig:fe_aa_dos}
\end{figure}

As mentioned in the discussion of equation~\ref{gf_c_dens}, the Green's function formalism allows us to change our integrals from being along the real-energy axis into contour integrals that extend into the upper-half of the complex plane \cite{starrett2015,gill17,starrett2018,Gill_2023}. Figure~\ref{fig:fe_aa_dos} shows the density of states (DOS) associated with the AA electronic structure for an iron plasma at 120~eV and 0.5~$\textrm{g/cm}^{3}$ along the top of a flat, complex contour (i.e. at a fixed imaginary part of the complex energy, $z=E+i\gamma$). The DOS (and all other integrands of interest) become broader the further the contour extends into the imaginary part of the complex energy plane. The broadening of sharp features includes the bound states at negative energies, which are infinitesimally narrow in energy along the real energy axis. The continuum edge is marked by a red vertical line, and beyond this edge we still see the presence of strong, narrow lines, which are the so-called continuum resonances. Integration over all three curves in figure~\ref{fig:fe_aa_dos}, along with appropriate closure of the contours, gives the same value, so we utilize the curve furthest off of the real-energy axis due to its smoothness and the ability to numerically resolve such a curve with very few grid points.

\begin{figure}[h]
    \centering
    \includegraphics[scale=0.33]{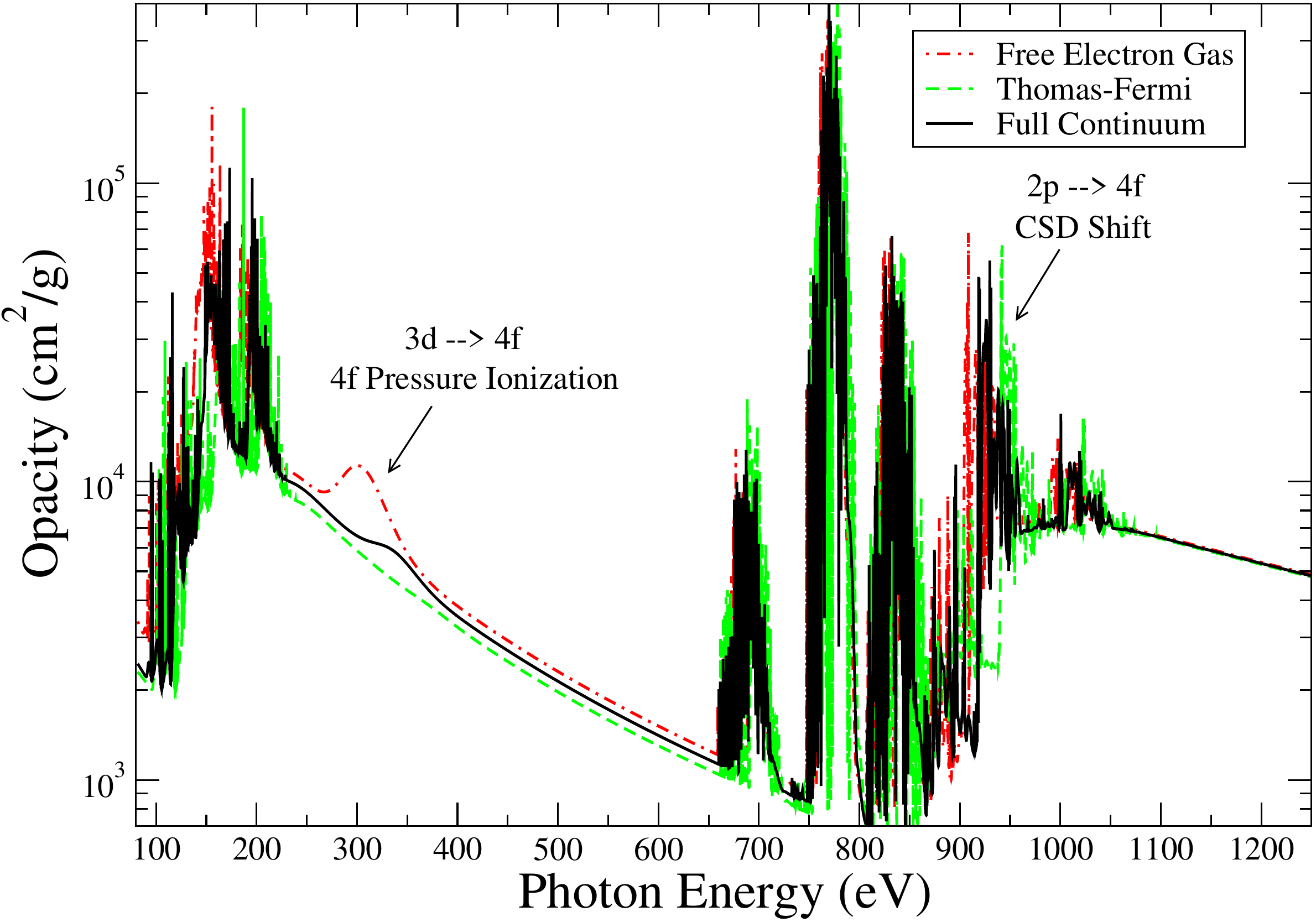}
    \caption[\fontsize{10pt}{10pt}]{The opacity spectrum of iron at 120~eV and 3.6~$\textrm{g/cm}^{3}$ is shown for models using different treatments of the continuum electrons within the SC structure calculations. The red dash-dot curve is produced from SCs whose continuum was represented by a homogeneous electron gas, which leads to the smallest plasma screening of the bound states. The green dashed curve is produced from SCs whose continuum was represented by a Thomas-Fermi approximation, which semi-classically describes the continuum and overpredicts the plasma screening. The black solid curve is produced with the full continuum model, which has a consistent treatment of bound and continuum electrons in the electronic structure. The location of the $3d \rightarrow 4f$ transitions is marked to indicate the impact of the pressure ionization of $4f$ electrons on the opacity spectrum. The $2p \rightarrow 4d$ transitions are marked to indicate that the CSD shifts due to the different models lead to shifts in the most prominent features of that portion of the opacity.}
    \label{fig:fe_ctm_shift}
\end{figure}

The ability to capture and treat the bound states and continuum resonances on an equal footing with a robust, efficient complex contour scheme is how our implementation is able to represent hybrid bound-continuum supershells in a consistent framework with purely bound supershells. This approach allows us to calculate electronic structure for our SCs in a way that is stable near pressure ionization conditions and includes a consistent exchange and correlation treatment between all electrons in the system, including the continuum states. 
The impact of this consistent treatment can be seen in the different opacity spectra shown in figure~\ref{fig:fe_ctm_shift}, which shows the opacity of an iron plasma at 120~eV and 3.6~$\textrm{g/cm}^{3}$ calculated using three different models for the continuum electrons. The red dash-dot curve corresponds to using a homogeneous electron gas to represent the continuum electrons, and this represents the simplest and least physically accurate model. The free electron gas model is included to show the limiting case of having no polarization of the continuum electron density and therefore demonstrates the weakest screening of the bound electronic structure by the continuum electrons. The green dashed curve corresponds to using a semi-classical Thomas-Fermi (TF) approximation for the continuum states, and this model is the most commonly employed in SC and STA calculations. The TF model predicts a stronger screening of the bound states by the continuum. The full continuum treatment (shown in the solid black curve) utilizes a consistent framework for the continuum states and bound states, as was described in section~2 and reference~\cite{Gill_2023}. The data in figure~\ref{fig:fe_ctm_shift} was previously shown in reference~\cite{Gill_2023}, and here we expand the discussion on the physical and model consequences that lead to the apparent discrepancies between the models.

When the full continuum model is applied, the $4f$ orbitals of some SCs do not exist as negative energy bound states due to plasma density effects, and these states instead appear as sharp resonances in the SC DOS. When the weakly screening free electron gas model is applied, the plasma density effects that are represented by the continuum electron screening are reduced, and therefore the $4f$ orbitals exist as bound states for all SCs. The presence of these orbitals throughout all SCs can be seen in figure~\ref{fig:fe_ctm_shift} where the $3d \rightarrow 4f$ bound-bound transitions are prominent. In the TF model, the stronger screening leads to the delocalization of the $4f$ orbitals for all SCs, and therefore the $3d \rightarrow 4f$ transitions are absent in the spectrum. 

\begin{figure}[h]
    \centering
    \includegraphics[scale=0.33]{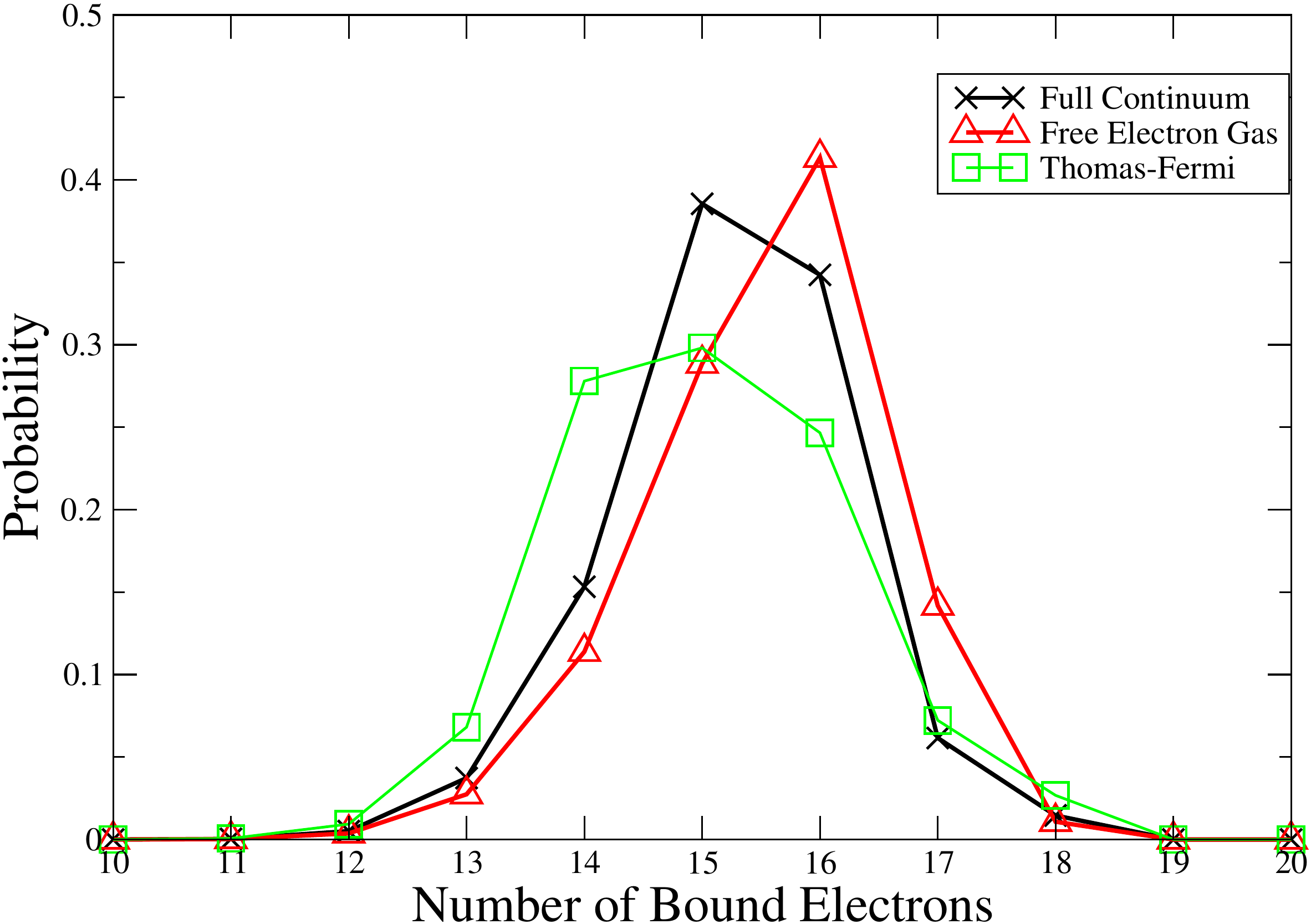}
    \caption[\fontsize{10pt}{10pt}]{The charge state distributions of an iron plasma at 120~eV and 3.6~$\textrm{g/cm}^{3}$ are shown. The different curves correspond to the CSDs associated with the opacities shown in figure~\ref{fig:fe_ctm_shift}.}
    \label{fig:fe_csd}
\end{figure}

The other significant consequence of using different treatments of the continuum electrons is seen in the charge state distributions (CSDs) of the plasmas. The prominent $2p \rightarrow 4d$ features marked on the right side of figure~\ref{fig:fe_ctm_shift} differ in each model mainly due to the fact that the differences in electron structure between the models lead to a shift in the populations of the SCs and therefore a shift in the CSDs of the plasma models. Figure~\ref{fig:fe_csd} shows the CSDs for the plasma models used to calculate the opacities of figure~\ref{fig:fe_ctm_shift}. It is clear that the weaker screening of the free electron gas model leads to a less-ionized plasma, while the TF model overpredicts the ionization and leads to a broader CSD. This shift in the CSD is translated to the SC populations used to calculate the strength of the opacity lines, and we therefore see the shift in bound-bound features marked in figure~\ref{fig:fe_ctm_shift}.

Though this work indicates that a consistent treatment of bound and continuum electrons is important when calculating electronic structure and the resulting opacity for dense plasmas, there is still unaddressed physics in these calculations. It should be noted that our SC populations are determined using the partition functions associated with only the bound electrons. This choice is made to provide consistency between our calculations and traditional isolated atom calculations in the low-density limit. However, the free energy of the continuum states should be incorporated into the SC partition functions in order to have a fully consistent bound and continuum framework within the plasma equation of state. This could be done in an approximate way by estimating the free energy of the continuum electrons using standard non-interacting fermion methods \cite{PAIN2006_ctmsta}, but such an approximate method would miss the electron-electron correlations that are included in the more sophisticated partition function approximations used to describe bound states \cite{gilleron2004,wilson2007_part,pain2020_recursion,pain2021}. 

Finally, the correlations between bound electrons are approximated using Slater integrals in order to obtain better estimations of the average energy associated with each SC \cite{cowan1981theory,Bar-Shalom89,bar-shalom96}. These Slater integrals play a crucial role in improving the STA energies of transitions, but all formulations of the STA moments include only the correlations between bound electrons. For example, the energies of the bound-bound transitions within a given ion-stage should approach the bound-free edge associated with the active orbital. However, the spectra in figure~\ref{fig:fe_ctm_shift} show the $3d\rightarrow 4f$ transitions at higher energies than the apparent $3d$ bound-free edge. This physically inaccurate, discontinuous behavior in the energies is due to the lack of a consistent treatment of correlations between the bound-bound and bound-free transition energies, and such a discrepancy has been shown to disappear when using the inaccurate (but consistent) independent particle approximations for transition energies \cite{Gill_2023}. STA calculations traditionally approximate the bound-free edge by considering the Slater integrals between bound and continuum electrons to be zero, and while this may be a good approximation for low-density systems, the bound and continuum electrons in high-density plasmas have significant correlations, as is evidenced by the change in electronic structure in our calculations. 

The capabilities developed and implemented in the \texttt{STAG} code show that consistency between bound and continuum electrons within the electronic structure calculations and within the definitions of supershells can lead to differences in opacity spectra for plasmas at high density. However, it is unclear whether inclusion of a comparable bound and continuum consistency in the calculation of plasma equation of state and the electron correlations needed to accurately describe transition energies would reduce or exacerbate the differences shown in figure~\ref{fig:fe_ctm_shift}. Investigating these issues will be the focus of future work.

\section{Conclusions}
We have described and demonstrated the extension of our previously developed SC and STA code capabilities to include a fully relativistic mode within the \texttt{STAG} code. The comparisons between the relativistic and non-relativistic STA opacities show the expected increase in spectral resolution due to the explicit treatment of $j$-resolved orbitals in the relativistic framework. We expanded the discussion of previously shown work, indicating the importance of a consistent treatment of bound and continuum electrons in SC electronic structure calculations of dense plasmsas \cite{Gill_2023}. The following two points were noted: (1) The apparent shifts in CSD that are observed when employing different models of continuum electrons do not include a consistent treatment of the continuum free energy within the plasma equation of state; proper inclusion of this free energy may lead to significant changes in the populations used to construct opacity. (2) The corrections to STA transition energies involve only approximating correlations between bound electrons, which leads to apparent inaccuracies for bound-bound transition energies near bound-free edges. 

\section*{Acknowledgments}
We thank T. Nagayama for providing the experimental data for iron. This work was supported by LANL’s ASC PEM Atomic Physics Project. LANL is operated by Triad National Security, LLC, for the National Nuclear Security Administration of the U.S. Department of Energy under Contract No.~89233218NCA000001.

\bibliography{phys}

\end{document}